%% file: COLOUR_FerbHol_Attack.tex
\documentclass[vecphys]{svmult}

\usepackage{makeidx}         
\usepackage{graphicx}        
\usepackage{multicol}        
\usepackage[bottom]{footmisc}
\usepackage{amssymb}
\usepackage{cite}

\makeindex             

\begin{document}

\title*{Attack Vulnerability of Public Transport Networks}
\author{Christian von Ferber\inst{1}$^,$\inst{2}, Taras
Holovatch\inst{3},\and Yurij Holovatch\inst{4}$^,$\inst{5}}
\authorrunning{C. von Ferber, T. Holovatch, Yu. Holovatch}
\institute{Applied Mathematics Research Centre, Coventry University,
Coventry CV1 5FB, UK \texttt{C.vonFerber@coventry.ac.uk}
 \and
 Physikalisches Institut, Universit\"at Freiburg, 79104 Freiburg,
Germany
 \and
 Ivan Franko National University of Lviv, 79005 Lviv,
Ukraine
 \and
 Institute for Condensed Matter Physics of the National Academy of Sciences of Ukraine,
 79011 Lviv, Ukraine
 \and
 Institut f\"ur Theoretische Physik, Johannes Kepler Universit\"at Linz, 4040 Linz, Austria
 \texttt{hol@icmp.lviv.ua}}
%
%

\maketitle

\abstract{The behavior of complex networks under attack
depends strongly on the specific attack scenario.
Of special interest are scale-free networks, which are
usually seen as robust
under random failure or attack but appear to be especially
vulnerable to targeted attacks. In
a recent study of public transport networks of 14 major cities
of the world we have shown that these networks may exhibit
scale-free behaviour [Physica A {\bf 380}, 585 (2007)]. Our
further analysis, subject of this report,
focuses on the effects that defunct or removed nodes have on
the properties of public transport networks.
Simulating different attack strategies we
elaborate vulnerability criteria that allow to find minimal
strategies with high impact on these systems.}

\section{Introduction}
\label{sec:1}

A number of different phenomena related to complex networks
\cite{netw_general} may be described in terms of
percolation theory\cite{perc_general}.
Take for example a network built following
given construction rules. Then, how should the rules be tuned
such that an infinite connected component is constructed with finite
probability and what are the properties of this class of networks when
the parameters reach the corresponding percolation threshold?
Taken that percolation is in general seen as a critical phenomenon
one may expect to find power laws in the vicinity of this point.
The network (class) being described by more than one parameter,
there are also many scenarios to cross the threshold exhibiting
different behavior of the observables.
Related questions are: how do infections spread on a network
and are there optimal immunization strategies?
These and similar questions are best formulated within percolation
theory \cite{perc_general} generalized from its original formulation
for regular grids to general network graphs.

In this paper we intend to apply concepts of complex network
theory  \cite{netw_general} to analyze the behaviour
of urban public transport networks (PTNs) under successive removal
of their constituents. In particular, continuing our recent study
of PTNs of 14 major cities of the world \cite{Ferber07a,Ferber07b},
we analyse their resilience against targeted
attacks following different scenarios.

It has been observed before that the behaviour of a complex network
under an attack that removes nodes or links may drastically differ
from that of regular lattices (i.e. from the classical percolation
problem). Early evidence of this fact was found analysing real world
scale-free networks: the www and the internet \cite{Albert00,Tu00},
as well as metabolic \cite{Jeong00}, food web \cite{Sole01}, and
protein \cite{Jeong01} networks. In these studies, the interest was
in the robustness of these networks subject to the removal of their
nodes. It appeared that these networks display an unexpectedly high
degree of robustness under random failure. However, if the scenario
is changed towards ``targeted'' attacks, the same networks may
appear to be especially vulnerable \cite{Cohen00,Callaway00}.

To check the attack resilience of a network, different scenarios of
attacks have been proposed: e.g. a list of vertices ordered by
decreasing degree may prepared for the unperturbed network and the
attack successively removes vertices according to this original list
\cite{Barabasi99,Broder00}. In a slightly different scenario the
vertex degrees are recalculated and the list is reordered after each
removal step \cite{Albert00}. In initial studies only little
difference between these two scenarios were observed
\cite{Callaway00}, however further analysis showed
\cite{Girvan02,Holme02} that attacks according to recalculated lists
often turn out to be more harmful than the attack strategies based
on the initial list, suggesting that the network structure changes
as important vertices or edges are removed.  Other scenarios
consider attacks following an order imposed by different
`centralities' of the nodes, e.g. the so-called betweenness
centrality \cite{Holme02}. In
particular for the world-wide airport network, it has been shown
recently \cite{Guimera04,Guimera05} that
 nodes with higher betweenness play a more important role in
keeping the network connected than those with high degree.

As it turns out, the behavior under attack of different real-world
networks, even if they are scale-free differ considerably; e.g.
computer networks behave differently than collaboration networks,
see \cite{Holme02}. Therefore, it is important to investigate in how
far the behaviour under attack of different real-world networks is
consistent or shows strong variations. Below we present some results
of our analysis for the PTNs of 14 major cities of the world (see
Ref. \cite{Ferber07a} and chapter \cite{Ferber07b} of this
volume for a detailed description of the included PTNs). A more
complete survey  will be a subject of a separate publication
\cite{Ferber07c}.

\section{Observables and attack strategies}
\label{sec:2}

In the analysis presented below we consider the PTNs of the
following cities: Berlin (number of stations $N=2996$, number of
routes $M=218$), Dallas ($N=6571$, $M=131$), D\"usseldorf ($N=
1544$, $M= 124$),  Hamburg ($N=  8158$, $M= 708$),  Hong Kong ($N=
2117$, $M= 321$), Istanbul ($N=  4043$, $M= 414$), London ($N=
11012$, $M= 2005$),  Los Angeles ($N=  46244$, $M= 1893$),  Moscow
($N=  3755$, $M= 679$),  Paris ($N=  4003$, $M= 232$),  Rome ($N=
6315$, $M= 681$),  S\~ao Paolo  ($N=7223 $, $M= 998$),  Sydney  ($N=
2034$, $M= 596$),  Taipei  ($N= 5311$, $M= 389$). This sampling
includes cities from different continents, with different concepts
of planning and different history of the evolution and growth of the
city and its PTN. For the purpose of this paper let the PTN of a
given city be given by the routes offered in this network. Each
route services a given ordered list of stations. Representing the
PTN in terms of a graph, we apply the following mapping: each
station is represented by a node; any two nodes that are
successively serviced by at least one route are connected by a
single link. We note that there are several other ways to represent
a PTN as a graph \cite{Ferber07a,Ferber07b,Sen03,Sienkiewicz05a}.
The particular representation that we use here is referred to as a
$\mathbb{L}$-space in Refs.
\cite{Ferber07a,Ferber07b,Sienkiewicz05a}.

The importance of a node $i$ of a given network $\cal{N}$ may be
measured by calculating a number of graph theoretical indicators.
Besides the node degree $k_i$, which in our representation equals
the number of nearest neighbours $z_1(i)$ of a given node $i$,
different centralities of the node may be defined as follows (see
e.g. \cite{Brandes01}:
\begin{eqnarray}\label{eq:1}
\mbox{closeness centrality \hspace{3em}} &&
 C_C(i)=\frac{1}{\sum_{t\in \cal{N}} \ell(i,t)}, \\
\label{eq:2} \mbox{graph centrality \hspace{3em}} &&
 C_G(i) = \frac{1}{{\rm
max}_{t\in \cal{N}} \ell(i,t)}, \\
\label{eq:3} \mbox{stress centrality \hspace{3em}} &&
 C_S(i)= \sum_{s\neq i \neq t\in \cal{N}} \sigma_{st}(i), \\
\label{eq:4} \mbox{betweenness centrality \hspace{3em}} &&
 C_B(i)= \sum_{s\neq i \neq t\in \cal{N}}
 \frac{\sigma_{st}(n)}{\sigma_{st}}.
 \end{eqnarray}
In Eqs. (\ref{eq:1})--(\ref{eq:4}), $\ell(i,t)$ is the shortest-path
length between a pair of nodes $i,t$ that belong to a network
$\cal{N}$, $\sigma_{st}$ is the number of shortest paths between
two nodes $s,t\in \cal{N}$, and $\sigma_{st}(i)$ is the
number of shortest paths between nodes $s$ and $t$ that go through
the node $i$. When observing a network under attack we
will also record the next nearest neighbours $z_2(i)$ and
the clustering coefficient $C(i)$ of all remaining nodes $n$. The latter is
the ratio of the number of links $E_i$ between the $k_i$
nearest neighbours of $i$ and the maximal possible number of mutual links between
them:
\begin{equation} \label{eq:5}
C(i)  =\frac{2E_n}{k_i(k_i-1)}.
\end{equation}

Note that the mean values of all the above introduced quantities are
well-defined for a connected network $\cal{N}$. However, some
of the analysed PTNs consist of several disconnected
components even before any perturbation is applied.
Moreover, the number of components naturally increases when
nodes are removed. Therefore, we restrict averages of the observables
to the largest network component $GCC\subset \cal{N}$.
We will indicate these averages by an over-line.
Nevertheless, some of quantities are also well defined for the whole
network, the corresponding average will be denoted by angular
brackets. An example we note the inverse shortest path length:
\begin{equation}\label{eq:6}
\langle \ell^{-1} \rangle = \frac{2}{N(N-1)}\sum_{i>j}\ell^{-1}(i,j)
\end{equation}
where the summation spans over all $N$ sites of the (possibly disconnected)
network and defining $\ell^{-1}(i,j)=0$ if nodes $i,j$ are disconnected.
Note that in this case $\langle \ell \rangle$ is
obviously ill-defined.

In what follows,
we will pursue a number of different attack strategies or
selection rules and criteria to
remove the nodes (vertices). In particular, the scenarios
are the following. ``Random vertex'' (RV): vertices (nodes) are removed in random order.
``Random neigbour'' (RN): one by one, a randomly chosen  neighbour of a
randomly chosen node is removed. This scenario appears to be effective
for immunization problems \cite{Cohen03} and it is based on the
fact, that this way nodes with a high number of neighbors will be
selected with higher probability.
In further scenarios nodes are removed according to the lists prepared in the
order of decreasing node degrees ($k$), centralities ($C(C)$,
$C(G)$, $C(S)$, $C(B)$), the number of their second nearest neighbours ($z_2$),
and increasing clustering coefficient ($C$). The latter seven
scenarios can be either implemented according to lists prepared for the
initial PTN before the attacks (we indicate the corresponding
scenario by a subscript $i$, e.g. $C_i(C)$) or the list is built by
recalculating the order of the remaining nodes after each step.
This way we follow sixteen
different strategies in attacking the networks. The observed changes
of the properties of the PTN
under these attacks are described in
the next section.

\section{Numerical results}
\label{sec:3}
\begin{figure}
\centering
\includegraphics[width=55mm]{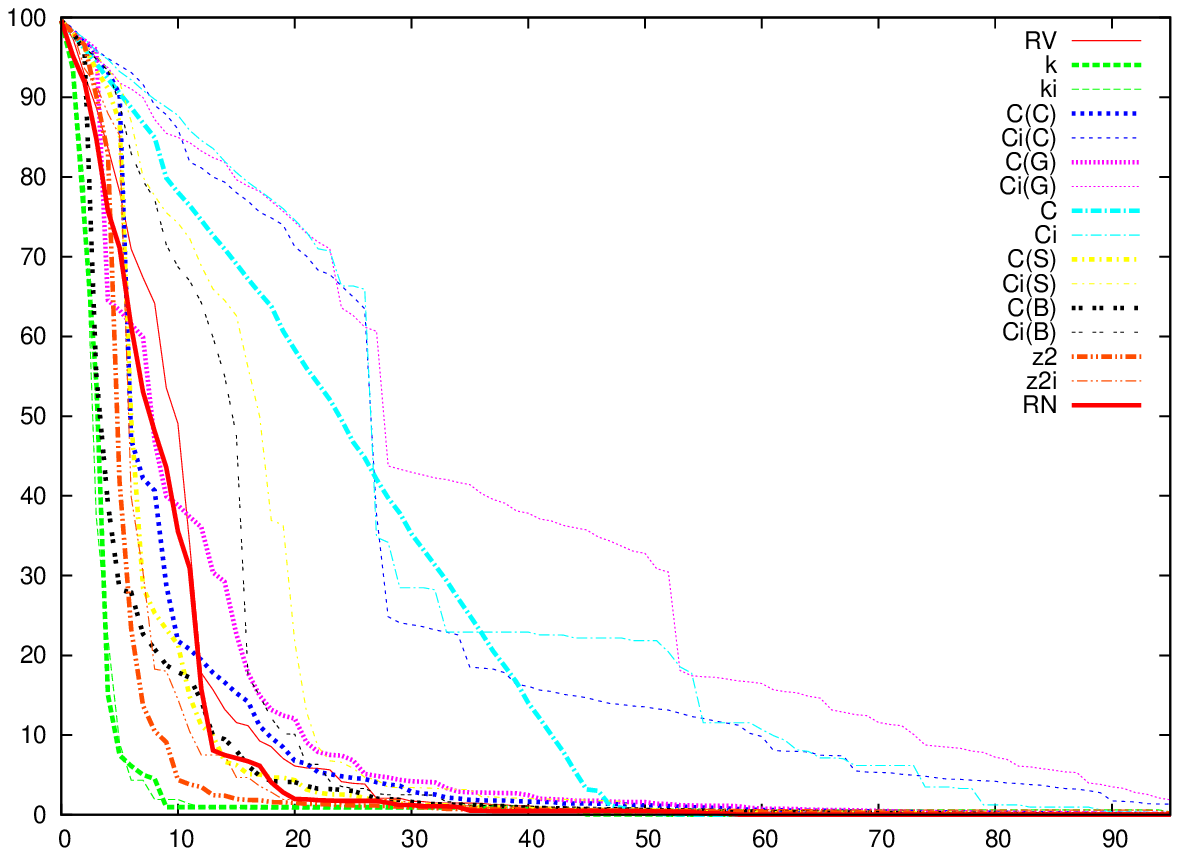}
\includegraphics[width=55mm]{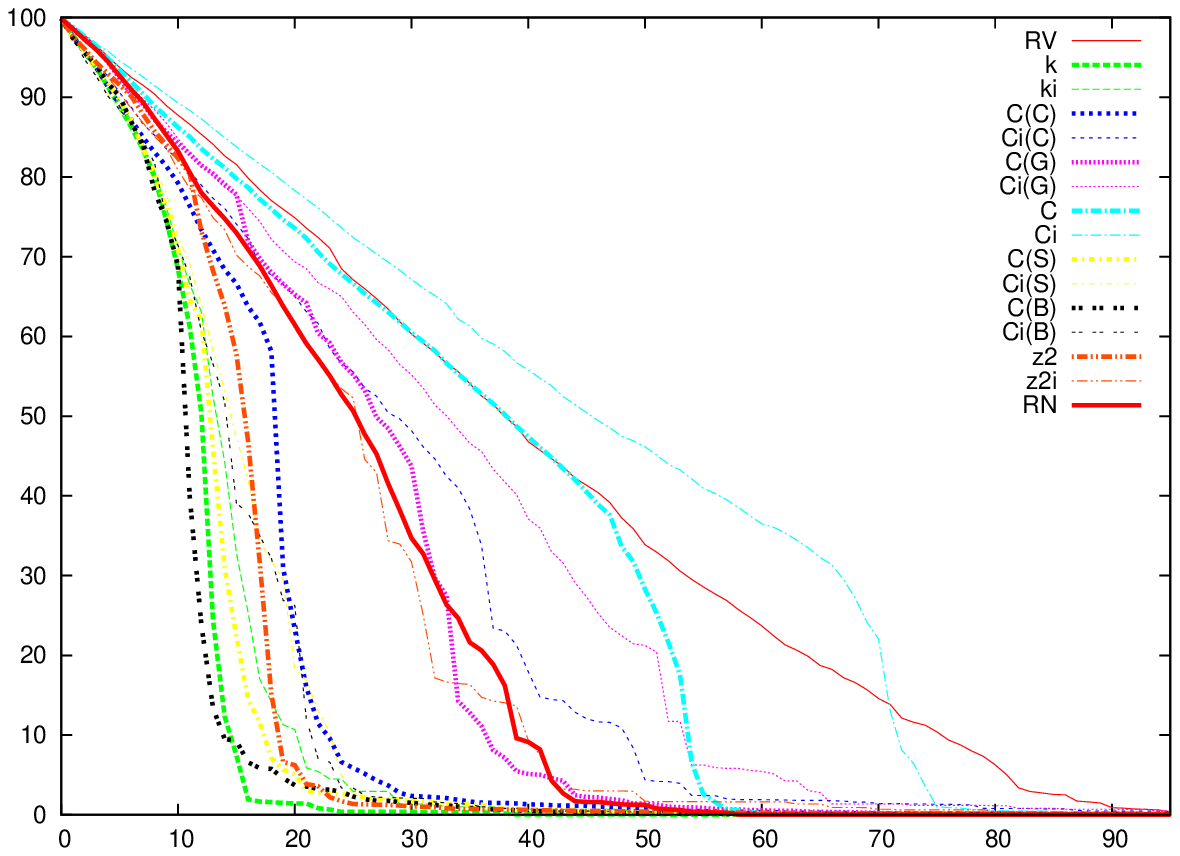}
{\bf{a)} \hspace{23em} \bf{b)}}
 \caption{Attacks on PTNs of (a) Dallas and (b) Paris.
Each curve corresponds to a different attack scenario
as indicated in the legend, see text. Horizontal axis: percents of removed nodes,
Vertical axis: normalized size $S$ of the largest component.}
 \label{fig:1}
\end{figure}

\begin{figure}
\centering
\includegraphics[width=55mm]{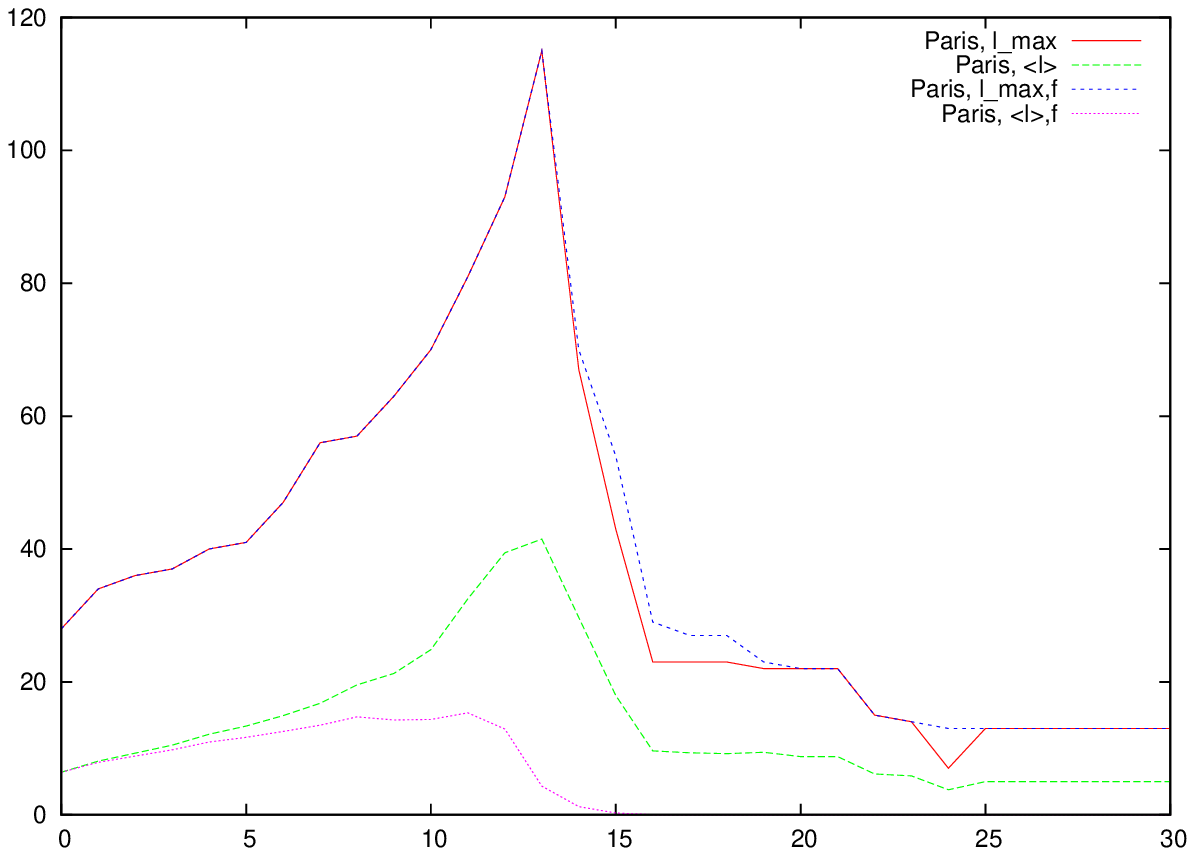}
\includegraphics[width=55mm]{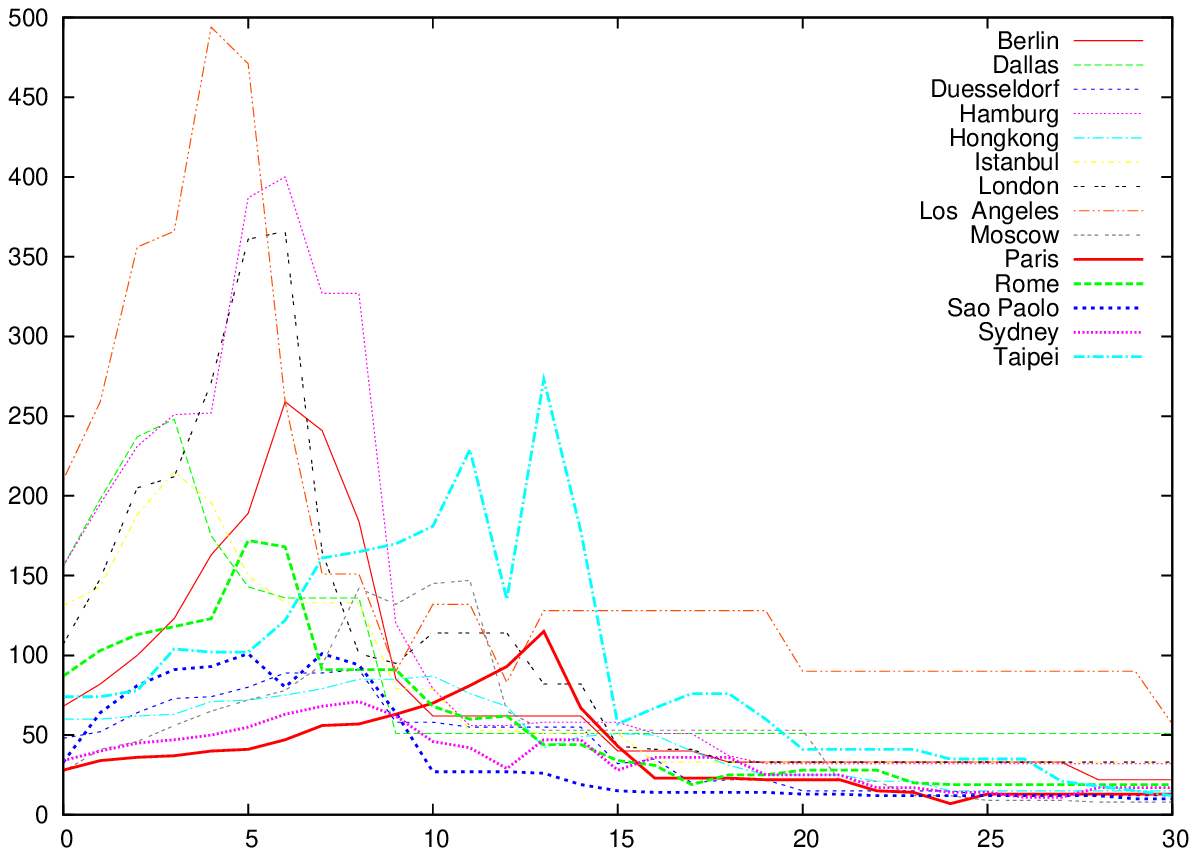}

{\bf{a)} \hspace{23em} \bf{b)}}
 \caption{Highest degree scenario. Horizontal axis as in Fig. \ref{fig:1}.
(a) Behavior of the maximal and mean shortest path lengths
for the Paris
PTN calculated for the largest component ($\ell_{\rm max}$,
$\overline{\ell}$) and  for the whole network ($\ell_{\rm max,f}$,
$<\ell>_{\rm f}$). Note a sharp maximum
occurs at 13 \% of removed nodes (stations) for $\ell_{\rm max}$,
$\overline{\ell}$, $\ell_{\rm max,f}$. (b) Behavior of the maximal shortest path
length $\ell_{\rm max}$ for the PTNs of different cities.
 \label{fig:3}}
\end{figure}

The theory of complex networks is concerned with the properties of
ensembles of networks (graphs) that are characterized e.g. by common
construction rules.
Such an ensemble is said to be in the percolation regime if even the
infinite graphs in this ensemble have a connected component that contains
a finite fraction of their nodes.
This component is referred to as the giant connected component GCC.
If the ensemble properties are controlled by some parameter, e.g.
the concentration of active nodes, then the percolation threshold
in terms of this parameter is defined as the value at which the network
ensemble enters the percolation regime.
In the present case of finite networks we denote by GCC the largest
connected component of a given network. For the finite networks
defined by the PTN we analyze the  behaviour
of the their largest component that contains  $N_{\rm GCC}$
nodes. We introduce the normalized largest component size $S$
by:
\begin{equation}\label{eq:7}
S = \frac{N_{\rm GCC}}{N} \times 100 \%.
\end{equation}
In Fig. \ref{fig:1} we show the behavior of $S$ for the
 attack strategies described above
for the PTNs of  Dallas and Paris.
At each step of the attack 1\% of the nodes is successively removed following
the selection criteria of the given scenarios. The effectiveness
of
the attack scenarios may be judged by their impact on the value of $S$.
As it is clearly seen from  Fig. \ref{fig:1}, the least effective
is the scenario of removing random nodes (RV): it is characterized
by the slowest decrease of $S$. Another obvious conclusion is that
scenarios based on lists calculated for the initial network
(marked by a subscript $i$) appear to be less harmful than those,
that are based on recalculated lists. Note however that the
difference between 'initial' and 'recalculated' scenarios is less
evident in the strategies based on the local characteristics, as
e.g. the node degree or the number of second nearest neighbours (c.f.
curves for $k$, $k_i$ and $z_2$, $z_{2i}$, respectively). The
above difference is even more pronounced for the centrality-based
scenarios. A principal difference between attacks on the highest
degree nodes on the one hand, and on the highest betweenness nodes
on the other hand is that the first quantity is a local, i.e. is
calculated from properties of the immediate environment of each node,
whereas the second one is global. Moreover, the first strategy aims to
remove a maximal number of edges whereas the second strategy
aims to cut as many shortest paths as possible.
Our analysis shows that the most effective are those scenarios
that are either targeted at nodes with the highest values of the node degree
$k$, the betweenness centrality $C_B$, the next nearest neighbour number
 $z_2$, or the stress centrality $C_S$ recalculated after each step of the
attack.
\begin{figure}
\centering
\includegraphics[width=55mm]{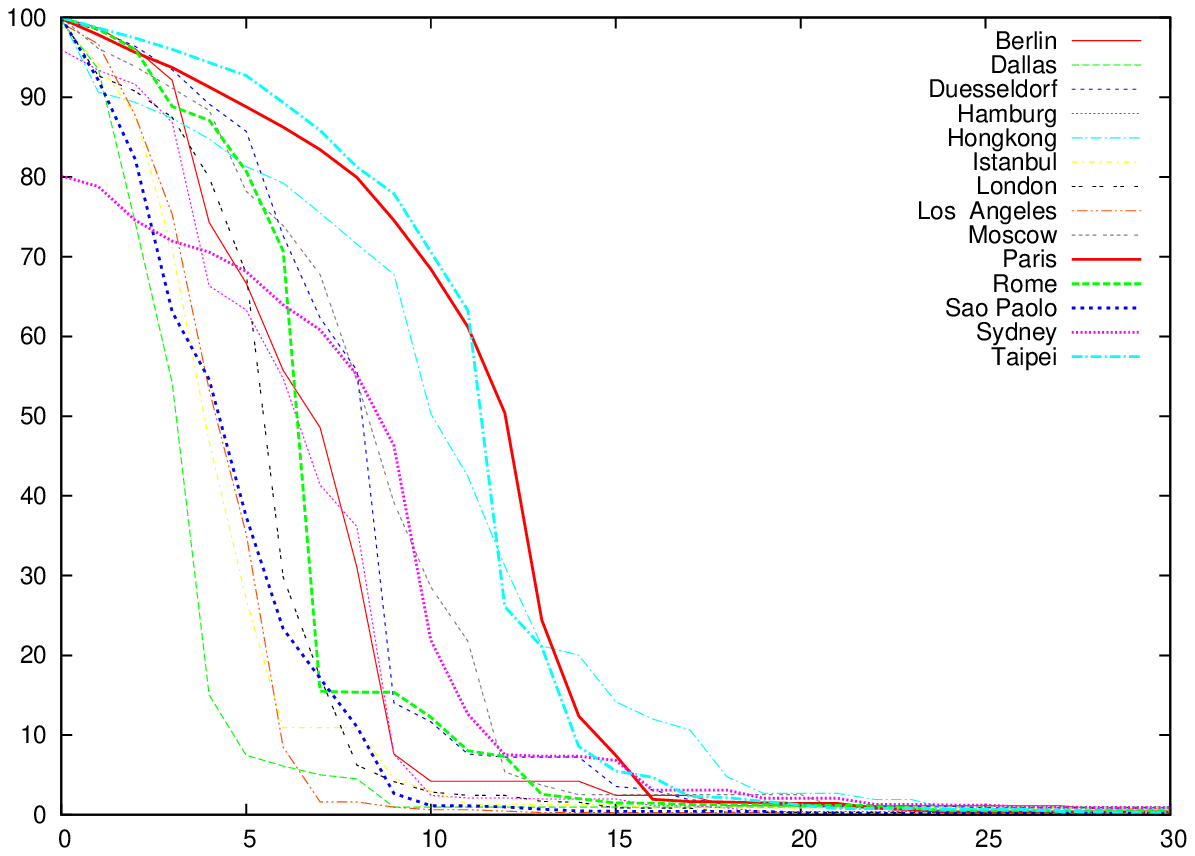}
\includegraphics[width=55mm]{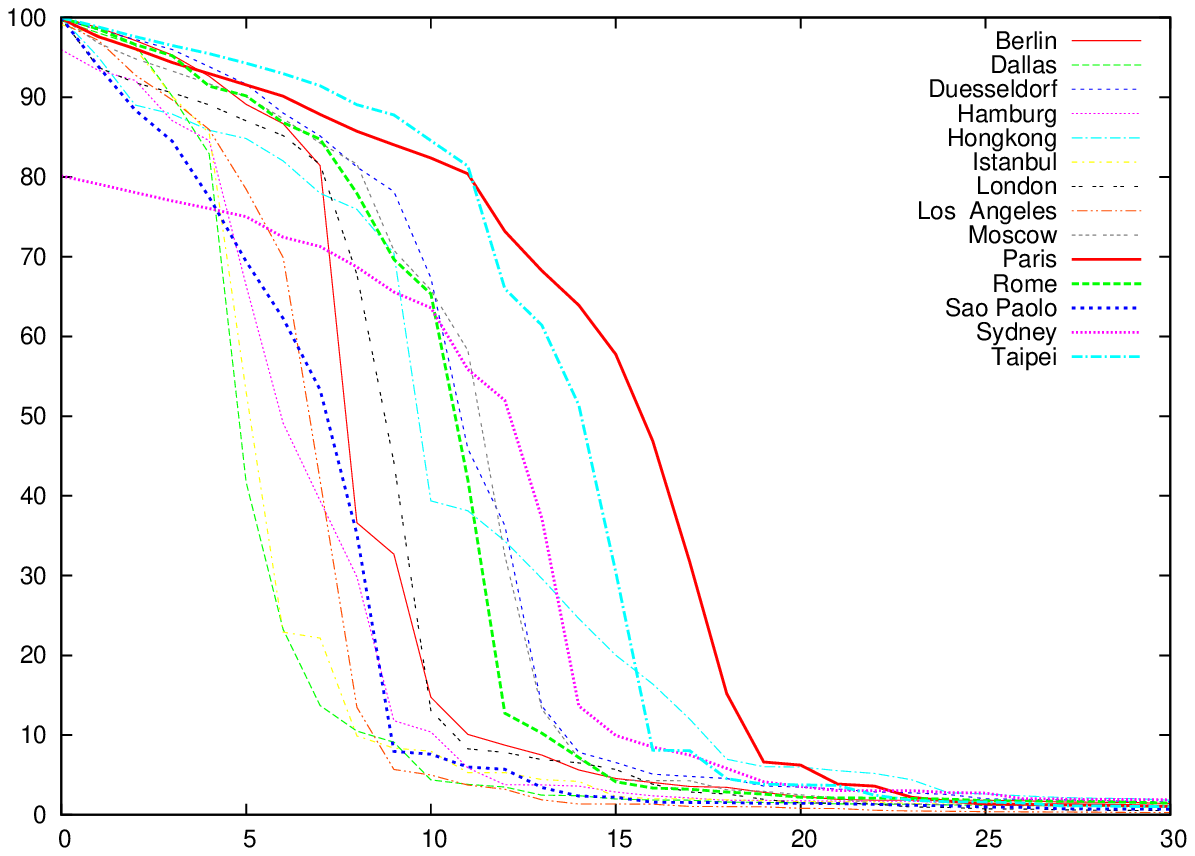}

{\bf{a)} \hspace{23em} \bf{b)}}

\includegraphics[width=55mm]{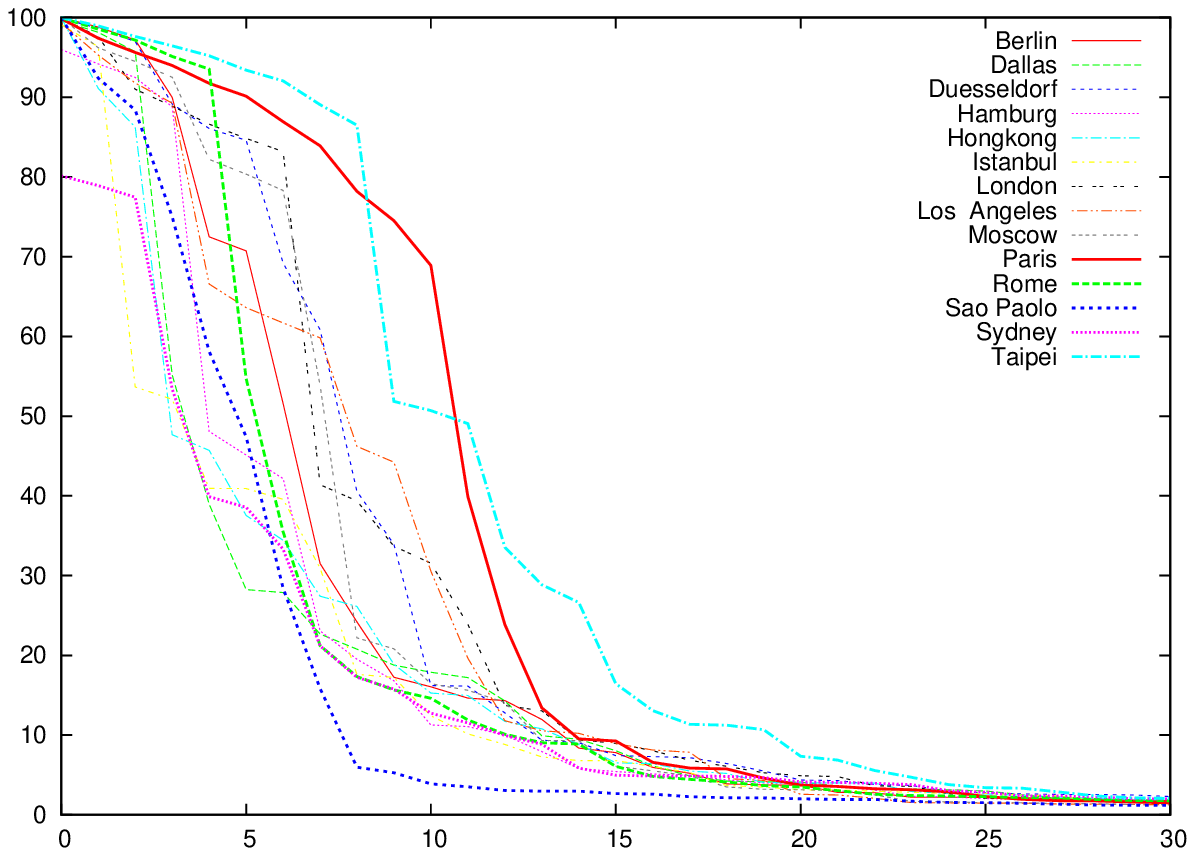}
\includegraphics[width=55mm]{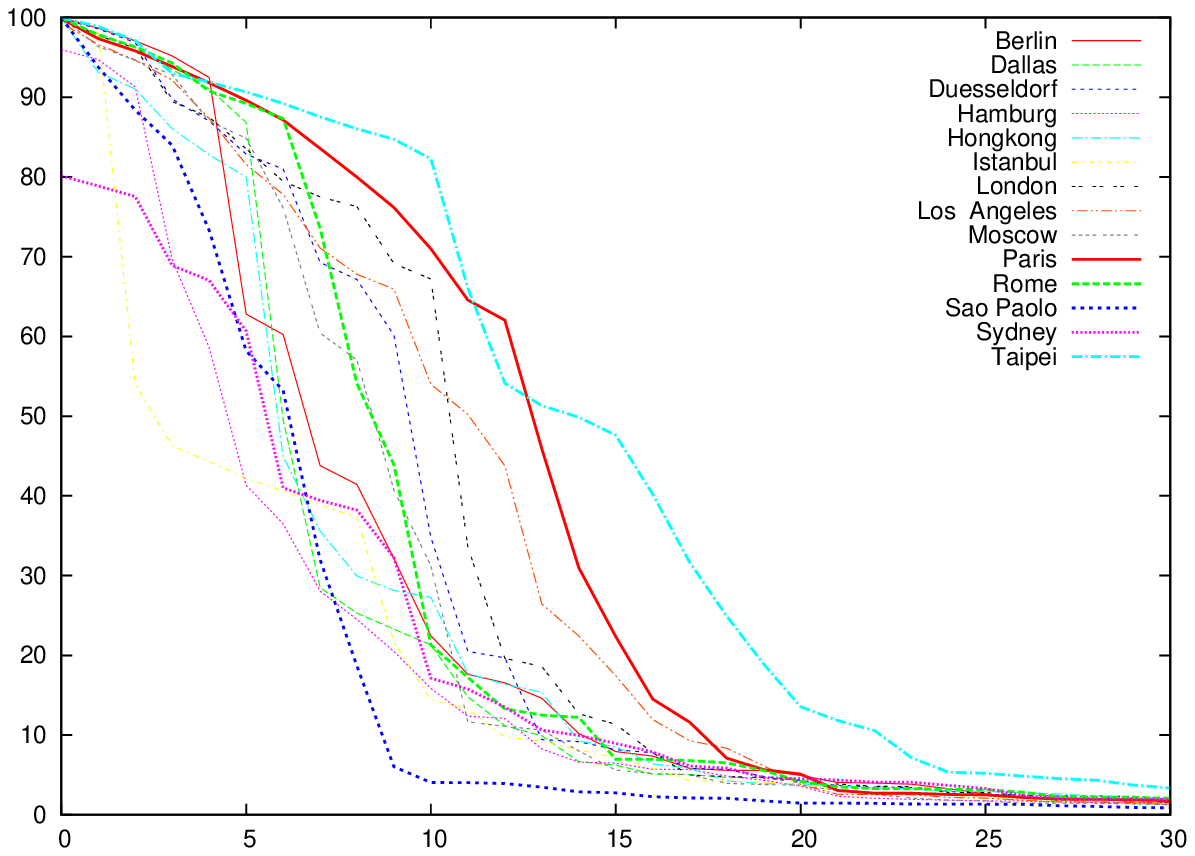}

{\bf{c)} \hspace{23em} \bf{d)}}
 \caption{Four attack scenarios for different PTNs (with recalculation):
attacks targeted at nodes of the highest (a) degree $k$,
(b) number of second neighbours $z_2$,
(c) betweenness centrality $C_B$, or  (d) stress centrality $C_S$.
Vertical and horizontal axis as in Fig. \ref{fig:1}.}
 \label{fig:2}
\end{figure}
 Figs.
\ref{fig:2}, \ref{fig:3} show that the order of destructiveness of
these scenarios differ for PTNs of different cities. However, among the
scenarios analyzed so far these four appear to be the most effective ones.

Another interesting quantity that we may deduce from Fig. \ref{fig:2}
is the vulnerability of the network in terms of the level of
destruction at which the largest network component breaks down. We
observe that this is strongly correlated to the initial value of the
so called Molloy-Reed parameter $\kappa=\overline{z}_2/\overline{z}_1$
of the unperturbed network.  Considering model networks that are
randomly built from sets of nodes with given degree distributions it
has been shown that the value of $\kappa_c=1$ represents the
percolation threshold in such networks \cite{Cohen03,Molloy}.  A value
much larger than $\kappa_c$ then indicates a significant distance from
the threshold.
The values of this parameter for the PTN studied here are:
Dallas ($\kappa=1.28$), Istanbul (1.54), Los Angeles (1.59),
Hamburg (1.85), London (1.87), Berlin (1.96), D\"usseldorf (1.96),
Rome (2.02), Sydney (2.54), Hongkong (3.24), S\~ao Paolo (4.17),
Paris (5.32), Moscow (6.24).
Comparing in particular with Fig. \ref{fig:2} a) we find indeed that
the higher the initial $\kappa$ value the less vulnerable the network
appears to be.

To more precisely define the threshold region for the concentration
of removed nodes we observe the behaviour of the  maximal
$\ell_{\rm max}$ and mean $\overline{\ell}$ shortest path lengths
under attack, as shown in Fig. \ref{fig:3}.
We focus on the recalculated degree scenario (k). Both maximal and
average path lengths display similar behaviour: initial growth and then an
abrupt decrease when a certain threshold is reached.Obviously,
removing the nodes initially increases the path lengths as
deviations from the original shortest paths need to be taken into account.
Further removing nodes then at some point leads to the breakup
of the network into smaller components on which the paths are naturally
limited by the boundaries which explains the sudden decrease of their
lengths. For the PTN of Paris we observe that this threshold is reached
for both $l_{\rm max}$ and $\overline{\ell}$ at
the same value of $c_{\rm segm}\simeq 13 \%$. The average shortest path
on all components of the network, $\langle\ell \rangle$, also
possesses a maximum in the same region (for the PTN of Paris it
occurs at $c\simeq 13 \%$).
However, the values of $c_{\rm segm}$ differ for
different cities (see Fig. \ref{fig:3},b)  and obviously strongly depend
on the attack scenario.

\begin{figure}
\centering
\includegraphics[width=55mm]{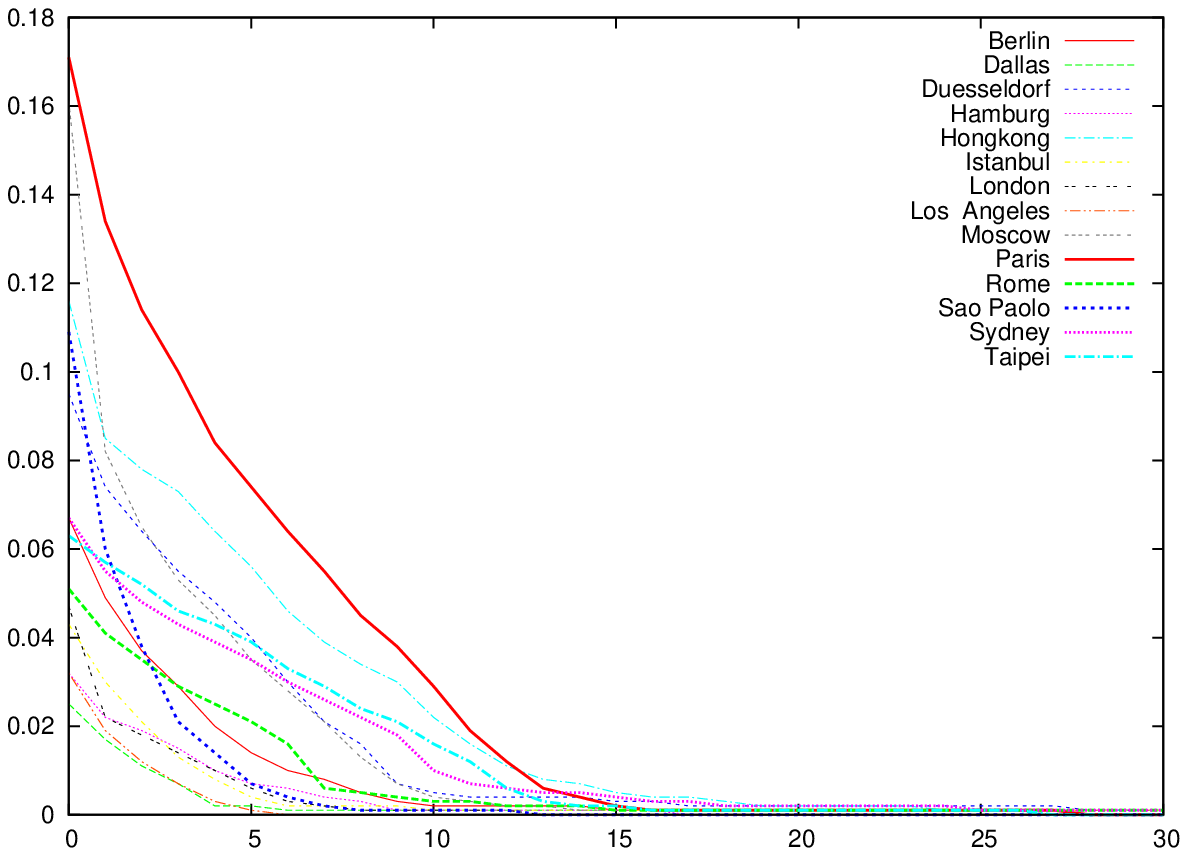}
\includegraphics[width=55mm]{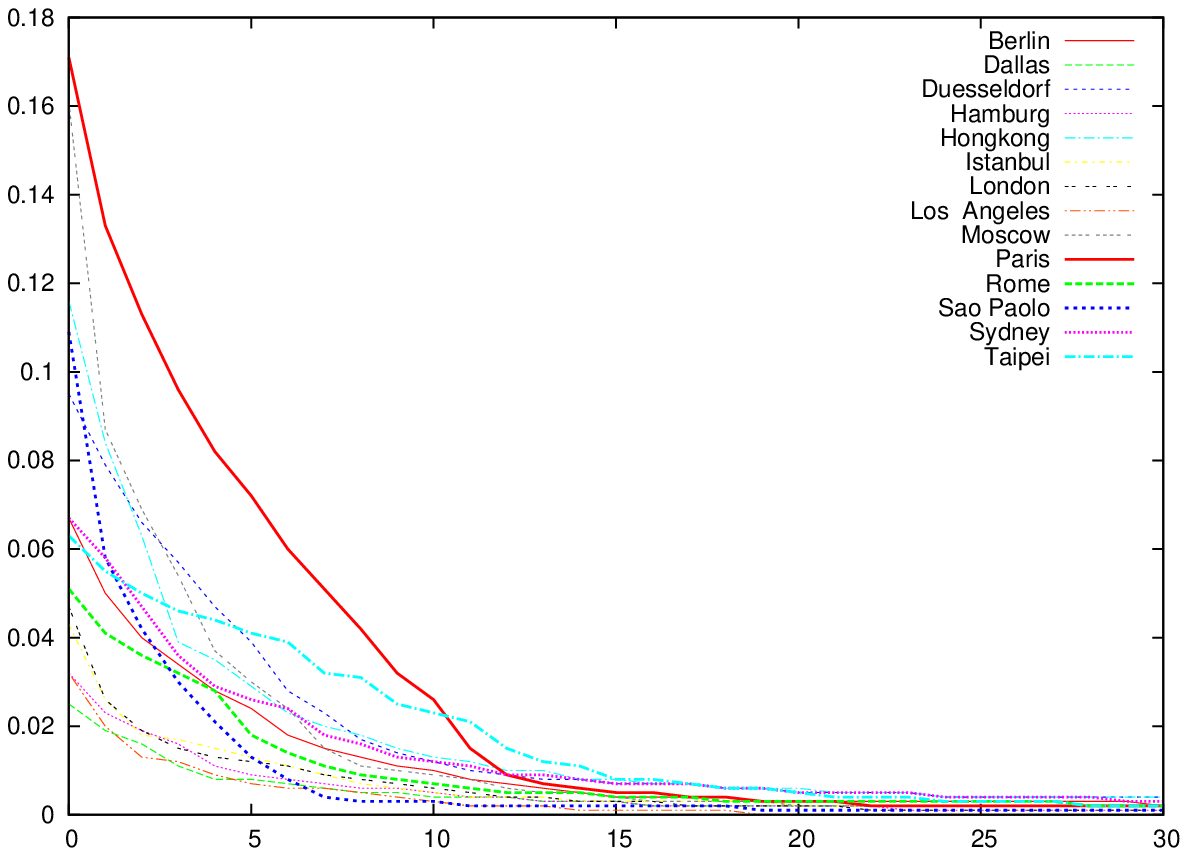}

{\bf{a)} \hspace{23em} \bf{b)}}
\includegraphics[width=55mm]{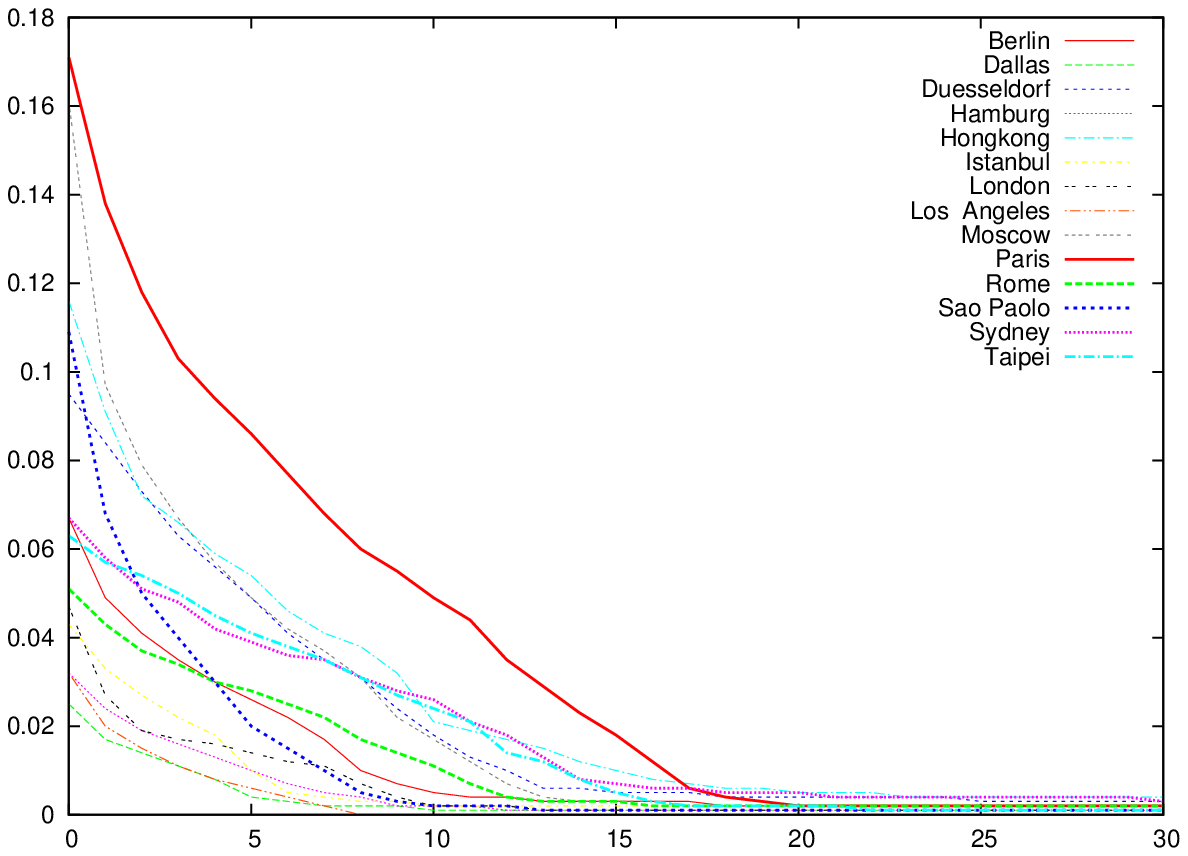}
\includegraphics[width=55mm]{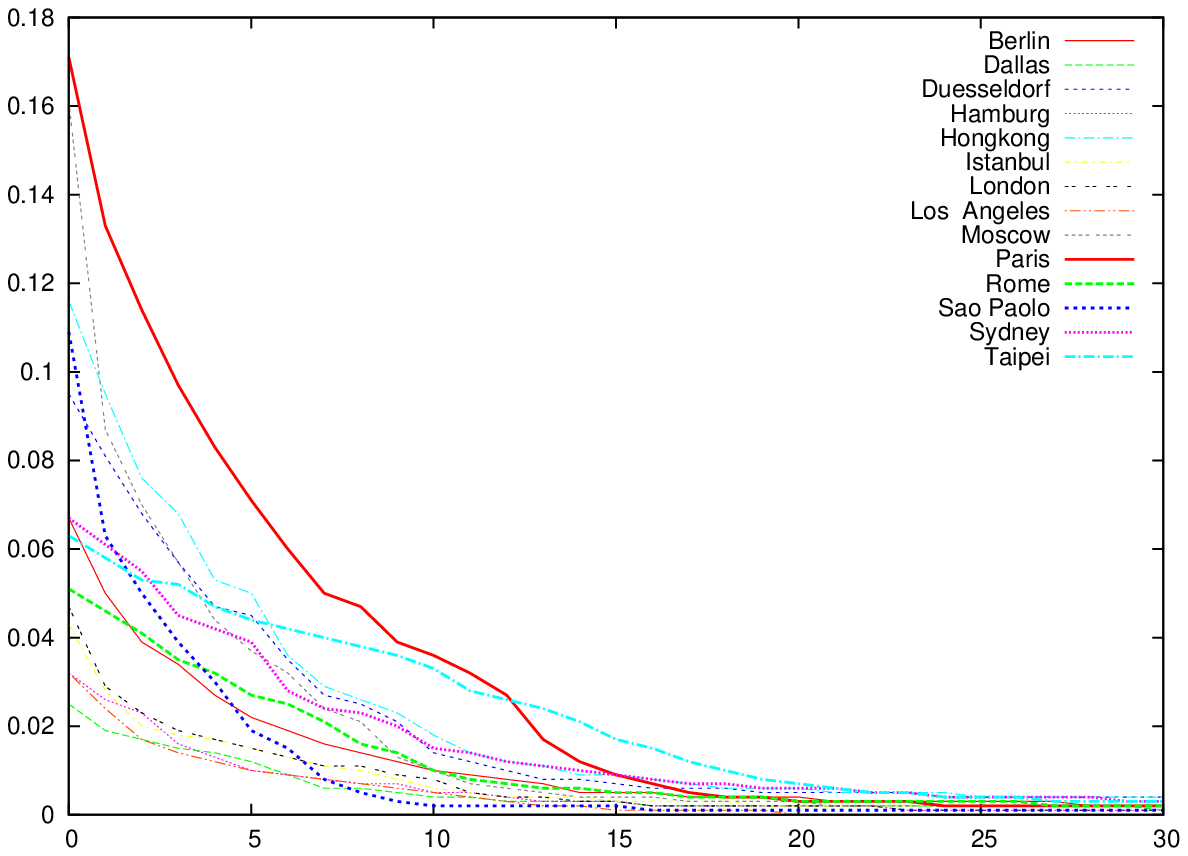}

{\bf{c)} \hspace{23em} \bf{d)}}
 \caption{Behaviour of $<\ell^{-l}>$ for PTNs of different cities
under attack following four different scenarios, see text:
a) highest $k$, b) highest $C_B$, c) highest $z_2$, d) highest $C_S$.
Horizontal axis as in Fig. \ref{fig:1}.}
 \label{fig:4}
\end{figure}

As discussed the observed maximum in $\ell_{\rm max}$ (or in
$\overline{\ell}$) appears to be a suitable criterion to identify
the values of $c$ (or at least the region in $c$), where the
segmentation of a network occurs. Other observables which resemble
an 'order parameter', are the above described largest connected
component size $S$, Eq.(\ref{eq:7}),  or the average value of the inverse
shortest path $\langle \ell^{-1} \rangle$ (\ref{eq:6}) are less
suitable for this purpose because of their rather smooth behaviour. In Fig.
\ref{fig:4}
we show for PTNs of fourteen cities the behavior of $<\ell^{-l}>$
under attacks following the four most harmful scenarios, i.e. the
recalculated highest $k$, $C_B$ $z_2$ and $C_S$ scenarios. Comparing
the impact of
 different attacks scenarios (as seen in particular in Fig.
\ref{fig:3}, \ref{fig:4}) one notices that the apparent relative impact
strongly depends on the choice of the observable
(e.g. $S$ or $<\ell^{-l}>$).

\begin{figure}
\centering
\includegraphics[width=38mm]{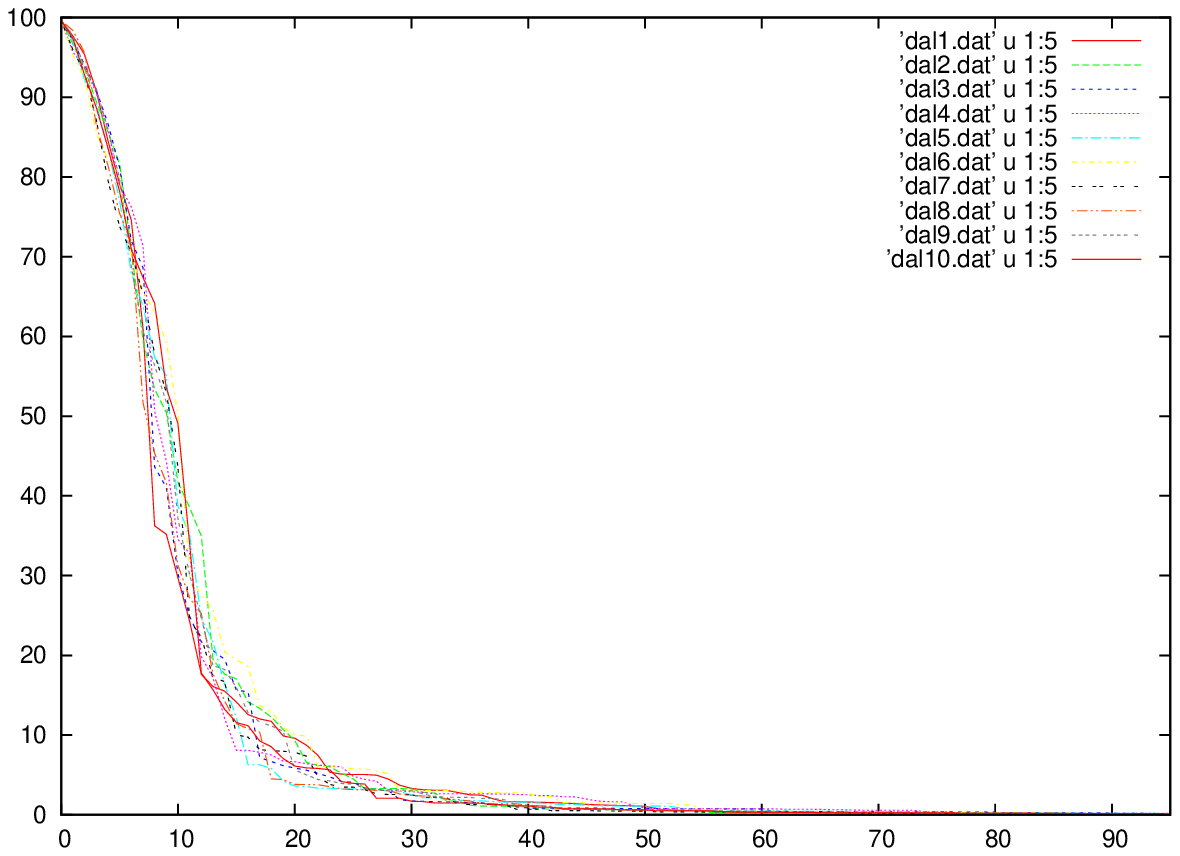}
\includegraphics[width=38mm]{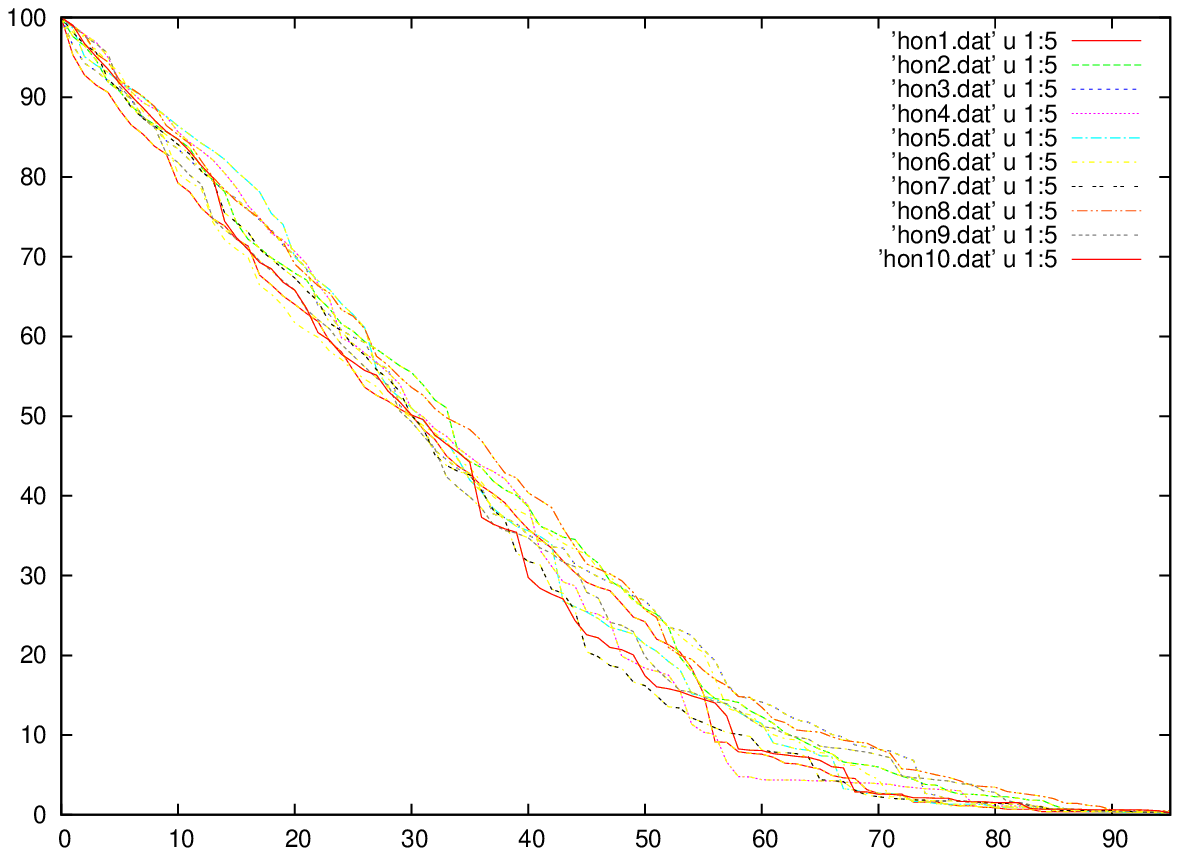}
\includegraphics[width=38mm]{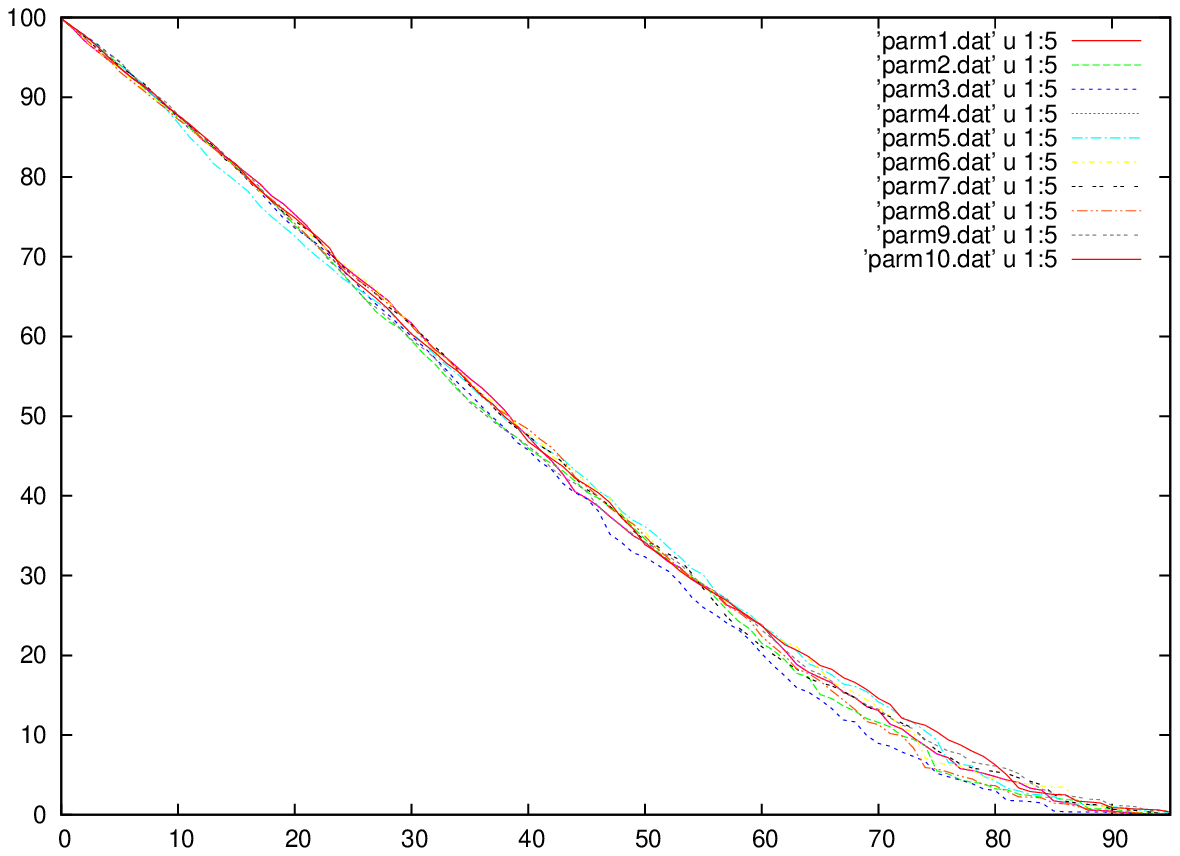}

{\bf{a)} \hspace{11em} \bf{b)} \hspace{11em} \bf{c)}}
 \caption{Impact and variance of the random vertex (RV) scenario on the
normalized size $S$  of the largest component for the PTNs of
(a) Dallas, (b) Hongkong, and (c) Paris.
Ten  curves of different colour indicate different
instances of the same scenario for each city.
Vertical and horizontal axis as in Fig. \ref{fig:1}}.
 \label{fig:5}
\end{figure}

\begin{figure}
\centering
\includegraphics[width=38mm]{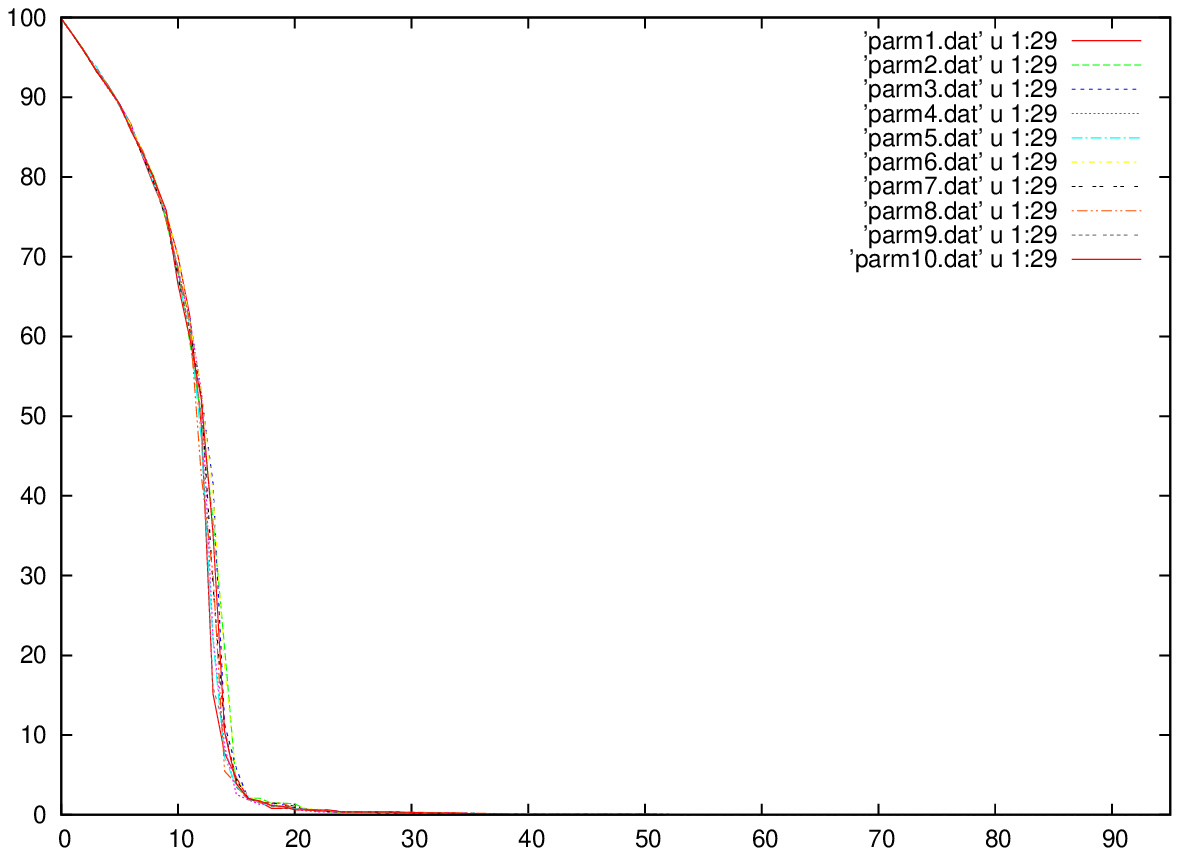}
\includegraphics[width=38mm]{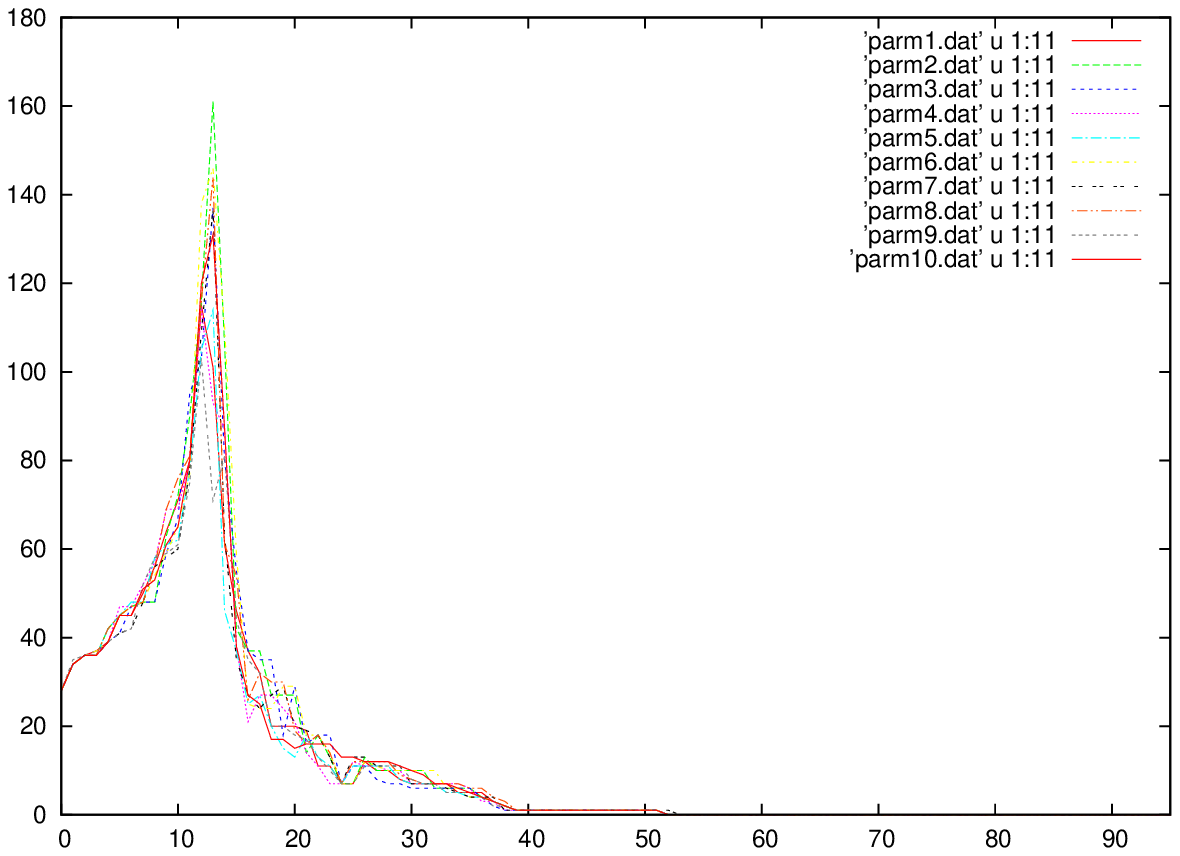}
\includegraphics[width=38mm]{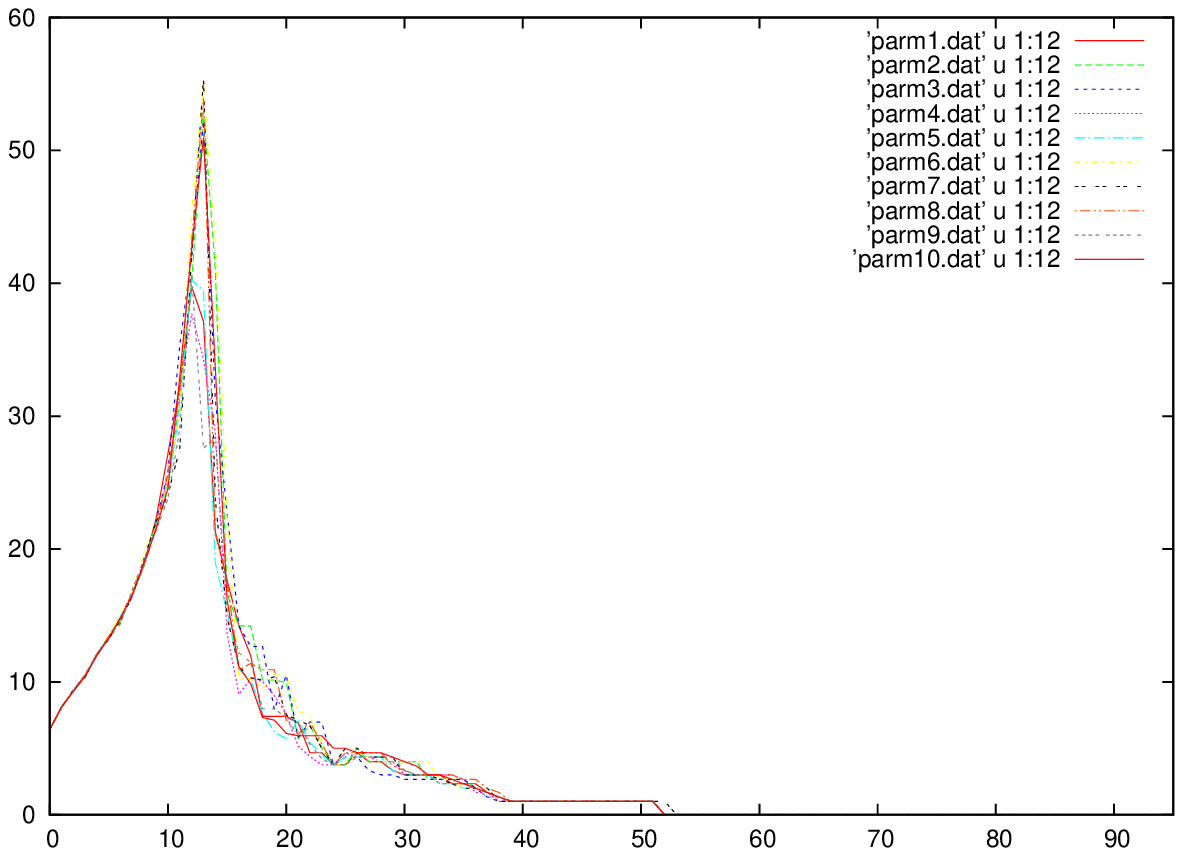}

{\bf{a)} \hspace{11em} \bf{b)} \hspace{11em} \bf{c)}}
 \caption{Ten instances of the recalculated highest degree scenario
for the PTN of Paris, observing: a) the largest connected component size $S$,
b) the maximal shortest path length $\ell_{\rm max}$, c) the mean shortest
path length $\overline{\ell}$. Horizontal axis as in Fig. \ref{fig:1} }
 \label{fig:6}
\end{figure}

It is worth to note the statistical origin of the data exposed so
far. Different instances of the same scenario may differ to some extent.
This is obvious for the random RV or RN scenarios, where the nodes are
removed according to a random procedure. However, it remains true even
for the attacks following pre-ordered lists of nodes. Obviously,
several nodes may have the same value with respect to a given
characteristic (e.g. $k$, $z_2$,or one of the centrality indices).
Then, the choice
between these nodes is random. To check the dispersion of the results,
Figs. \ref{fig:5}, \ref{fig:6} show the results of 10 complete
attack sequences for the same scenario. Fig. \ref{fig:5}
shows the change in the largest connected component $S$ of the PTNs
of Dallas (a), Hongkong (b), and Paris (c) for the random vertex (RV)
 scenario. The scatter of the curves in each figure provides an
idea about the deviations between individual samples.
The figures also clearly show that even attacked randomly, PTNs of
different cities may display a range of different behaviour: from the
comparatively fast decay of the largest connected component (as in
the case of Dallas, Fig. \ref{fig:5}a) to very slow, nearly linear decay
(as in the case of Paris, Fig. \ref{fig:5}c).

The dispersion in the largest connected component size $S$ is much
less for sequences of targeted attacks. In Fig. \ref{fig:6} we show
the behavior of the largest cluster size and the maximal and mean
shortest path lengths for the Paris PTN
for ten complete attack sequences following the recalculated degree (k)
scenario. Besides a rather narrow scattering of the data
for $S$ one notes, that within the current resolution the locations of
the maxima in $\ell_{\rm max}$ and $\overline{\ell}$ are very robust.

To give an idea for the numerical values of different
characteristics of the PTN as monitored during our analysis we display in Table
\ref{tab:1} some data for the PTN of Paris
for the recalculated degree scenario for some points of the sequence between
the unperturbed network and the vicinity of the threshold (maximum of
the shortest path lengths).
\begin{table}
\centering \caption{PTN of
Paris during an attack sequence following the recalculated degree
scenario. $c$: \% of
removed nodes; $N$: number of remaining nodes;
$\overline{k}=\overline{z}_1$: mean node degree;
$\overline{z}_2/\overline{z}_1$: ratio of the mean second to the mean
first nearest neighbour number;  $\ell_{\rm max}$: maximal shortest
path length; $\overline{\ell}$: mean shortest path length; $\langle
\ell^{-1} \rangle $: mean inverse shortest path length (for all of the
remaining network); $\overline{C}_C$, $\overline{C}_G$,
$\overline{C}_S$, $\overline{C}_B$: mean closeness, graph, stress,
and betweenness centralities; $\overline{C}$: mean clustering
coefficient; $S$: normalized largest component size.
\label{tab:1}} \tabcolsep=1mm
\begin{tabular}{lllllllllllll}
 \hline\noalign{\smallskip}
$c$ &   $N$  &    $\overline{k}=\overline{z}_1$ &
$\overline{z}_2/\overline{z}_1$ & $\ell_{\rm max}$ &
$\overline{\ell}$ & $\langle \ell^{-1} \rangle $ & $\overline{C}_C$
& $\overline{C}_G$ & $\overline{C}_S$ & $\overline{C}_B$ &
$\overline{C}$ & $S$ \\
 \noalign{\smallskip}\hline\noalign{\smallskip}
0   &   3728 &      3.73 &    5.32 &      28 &    6.41 & 0.17 &
0.004    &    5.47 &   38167 &   10062 & 0.079 &
99.87\\
1   &   3691  &     3.25 &    3.40 &      34 &    8.08  &      0.13  &      0.003    &    4.64 &   40419 &   12912 &        0.073 &    97.85\\
5   &   3543  &     2.52 &    2.05 &      41 &   13.35  &      0.07  &      0.002    &    3.60 &   50496 &   20439 &        0.062 &    88.81\\
10  &    3358 &     2.00 &    1.43 &      70 &   24.84  &      0.03  &      0.002    &    2.02 &   53654 &   30406 &        0.044 &    68.45\\
12  &    3284 &     1.84 &    1.25 &      93 &   39.44  &      0.01  &      0.001    &    1.42 &   56218 &   36097 &        0.036 &    50.40\\
13  &    3247 &     1.77 &    1.19 &     115 &   41.49  &      0.01  &      0.003    &    1.21 &   31803 &   18404 &        0.039 &    24.41\\
14  &    3210 &     1.70 &    1.13 &      67 &   29.69  &      0.00  &      0.008    &    1.90 &   11915 &    6598 &        0.022 &    12.37\\
 \noalign{\smallskip}\hline
\end{tabular}
\end{table}
\section{Conclusions}
\label{sec:4}

In this paper we reported on some results concerning the behavior
of PTNs under attacks.
Similar to other real-world and model complex
networks \cite{Albert00,Tu00,Jeong00,Sole01,Jeong01,Holme02}, the
PTNs manifest very different behaviour under attacks of different
scenarios. With some notable exceptions
they appear to be robust to random attacks but more vulnerable
to attacks targeted at nodes with particular importance
as measured by the values of
certain characteristics (the most significant being the first and
second neighbour numbers, as well as the betweenness and stress centralities).
The observed difference between attack scenarios based on the initial and
the recalculated distributions shows that the network structure changes
essentially during the attack sequence. This is necessarily to be taken into
account in the construction of efficient strategies for the protection of
these network.

As a suitable criterion to identify the level of resilience, i.e.
the number of removed nodes
that leads to segmentation it is  useful to observe the
behaviour of the maximal shortest path length $\ell_{\rm max}$. For
the majority of PTNs networks we have analyzed here this observable
displays a sharp maximum as function of the removed node concentration
which indicates the breakup of the network. Other
'order-parameter-like' variables like the largest connected component
size $S$ or the average value of the inverse shortest path $\langle
\ell^{-1} \rangle$ are less suitable for this purpose because of
their smooth behaviour. Another observation is that in the recalculated
highest-degree attack scenario for the
segmentation often occurs at a value of
$\kappa=\overline{z}_2/\overline{z}_1\sim 1$ (see
e.g. Table \ref{tab:1} for Paris). Although the PTNs are correlated
structures, the above estimate resembles the Molloy-Reed
\cite{Molloy} criterion for randomly built uncorrelated networks.
Further investigation is needed to understand the mechanisms
that lead to higher resilience against random failure as observed
e.g. for the Paris network and how this behavior is related to the
network architecture.

As mentioned in the introduction, there are
different graph representations, also called `spaces',
for a given PTN
\cite{Ferber07a,Ferber07b,Sen03,Sienkiewicz05a}.
These will also lead to different connectivity relations and
path lengths between nodes. The resilience of PTNs in these
more general `spaces' will be discussed elsewhere \cite{Ferber07c}.

\section*{Acknowledgments}
Yu.H. acknowledges financial support of the Austrian Fonds zur
F\"orderung der wi\-ssen\-schaft\-li\-chen Forschung under Project
P19583. C.v.F. was supported in part by the EC under the Marie Curie Host
Fellowships for the Transfer of Knowledge MTKD-CT-2004-517186.

%
\input{COLOUR_FerbHol_Attack_ref}



\printindex
\end{document}

%% file: COLOUR_FerbHol_Attack_ref.tex
%
%

%
%